\documentclass[12pt]{iopart}

\usepackage{iopams}
\usepackage{verbatim} 
\usepackage[utf8]{inputenc}

\usepackage{graphicx}
\usepackage[colorlinks=true,citecolor=blue]{hyperref}
\usepackage{units}

\newcommand{\bra}[1]{\langle #1|}
\newcommand{\ket}[1]{|#1\rangle}
\renewcommand{\vec}[1]{\mathbf{#1}}

\def\up{\uparrow}
\def\down{\downarrow}
\def\kB {k_\mathsf{B}}

\begin{document}

\title{Non-equilibrium current and noise in inelastic tunnelling through a magnetic atom}

\author{Bj\"orn Sothmann and J\"urgen K\"onig}
\address{Theoretische Physik, Universit\"at Duisburg-Essen and CeNIDE, 47048 Duisburg, Germany}
\eads{\mailto{bjoerns@thp.uni-due.de},\mailto{koenig@thp.uni-due.de}}

\date{\today}

\begin{abstract}
In a recent experiment, Hirjibehedin {\it et al.} [Science {\bf 317}, 1199 (2007)] performed inelastic tunnelling spectroscopy of a single iron atom absorbed on a nonmagnetic substrate. The observed steps in the differential conductance marked the spin excitation energies. In this paper, we explain observed nonmonotonicities in the differential conductance by a nonequilibrium population of the atom spin states. Furthermore, we predict super-Poissonian current noise due to this nonequilibrium situation. We argue that the remarkable absence of nonequilibrium features at certain conductance steps indicates the presence of an anisotropic relaxation channel.
\end{abstract}

\pacs{68.37.Ef,73.23.Hk,72.70.+m,73.20.Hb}

\submitto{\NJP}

\maketitle
\section{Introduction}
Inelastic scattering processes provide a convenient tool to study the excitation spectra of various systems. By using inelastic electron tunnelling spectroscopy, one can access vibrational excitations of ensembles of molecules in metallic tunnel junctions~\cite{jaklevic_molecular_1966,scalapino_theory_1967} or single molecules in scanning tunnelling microscope (STM) geometries~\cite{stipe_single-molecule_1998}. The excitation energies reveal themselves as steps in the differential conductance whenever a new inelastic transport channel opens up. For the explanation of the measured signals an equilibrium distribution of the molecule states was  implicitly assumed. Further studies of molecular vibrations were performed using H$_2$ molecules~\cite{smit_measurement_2002}, C$_{60}$ molecules~\cite{park_nanomechanical_2000} in mechanical break junctions and suspended carbon nanotubes~\cite{leroy_electrical_2004,leturcq_franck-condon_2009,httel_pumping_2009}.

Recently, the investigation of magnetic properties and interactions on an atomic level became possible due to the advent of spin inelastic electron spectroscopy~\cite{heinrich_single-atom_2004,hirjibehedin_spin_2006,meier_revealing_2008,hirjibehedin_large_2007,otte_role_2008,otte_spin_2009}. Here, single magnetic atoms absorbed on a nonmagnetic substrate were contacted using an STM tip. Describing the atom in terms of a localized spin, Hirjibehedin {\it et~al.}~\cite{hirjibehedin_large_2007} related the positions of the conductance steps to the energy associated with transitions between different eigenstates, while the relative step heights depend on the matrix elements of the spin operator. A more complete theoretical description based on perturbation theory in the tunnel coupling \cite{fernndez-rossier_theory_2009,fransson_spin_2009,persson_theory_2009,lorente_efficient_2009,delgado_spin-transfer_2010} still relies on the assumption of equilibrium occupations.

While the above studies could explain the conductance steps assuming the atom spin to be in thermal equilibrium with the substrate, nonmonotonic features clearly present in the experimental results of~\cite{hirjibehedin_large_2007} were not addressed. Conductance overshoots due to nonequilibrium occupations together with their relaxation by spin-phonon interactions have been discussed in~\cite{lehmann_cotunneling_2006} for cotunnelling through a quantum dot. A similar behaviour was found in~\cite{rosch_nonequilibrium_2003,paaske_nonequilibrium_2004,schoeller_real-time_2009} who additionally studied the low-temperature nonequilibrium logarithmic Kondo enhancement of this overshoot. While the Kondo effect is relevant for transport through a single Co atom studied in~\cite{otte_role_2008,otte_spin_2009}, it is not important here. Nonequilibrium effects have been considered in~\cite{delgado_spin-transfer_2010} for spin-transfer torque on a single atom coupled to ferromagnetic substrates and tips. In~\cite{romeike_spin_2006}, the nonequilibrium current and current noise through a single molecular magnet was analyzed in the charge fluctuation regime. In this paper, we explain the experimental results by calculating the nonequilibrium occupations together with a spin-dependent relaxation channel using a master-equation approach. We, furthermore, predict an enhanced Fano factor indicating super-Poissonian current noise as a clear sign of a nonequilibrium situation.

\section{Model}
\begin{table}
	\caption{\label{tab:eigenstates}Eigenenergies $E_m$ and eigenstates $|m\rangle$ of the spin Hamiltonian~\eref{eq:spin} in the basis $|S_z\rangle_z$ of the $S_z$ eigenstates for a magnetic field applied in the $z$ and the $x$ direction, respectively.}
	\begin{indented}\lineup
	\item[]\begin{tabular}{@{}lllllll}
		\br
		$B_z=\unit[7]{T}$ & $E_m$(meV) & $|2\rangle_z$ & $|1\rangle_z$ & $|0\rangle_z$ & $|-1\rangle_z$ & $|-2\rangle_z$ \\
		\mr
		$\ket{0}$ & \-7.982 & 0.021 & 0 & \-0.097 & 0 & 0.995\\
		$\ket{1}$ & \-4.612 & 0.987 & 0 & \-0.157 & 0 & \-0.036\\
		$\ket{2}$ & \-2.813 & 0 & 0.402 & 0 & \-0.916 & 0\\
		$\ket{3}$ & \-0.287 & 0 & 0.916 & 0 & 0.402 & 0\\
		$\ket{4}$ & 0.194 & 0.159 & 0 & 0.983 & 0 & 0.092\\
		\br
	\end{tabular}
	\end{indented}
	\begin{indented}\lineup
	\item[]\begin{tabular}{@{}lllllll}
		\br
		$B_x=\unit[3]{T}$ & $E_m$(meV) & $|2\rangle_z$ & $|1\rangle_z$ & $|0\rangle_z$ & $|-1\rangle_z$ & $|-2\rangle_z$ \\
		\mr
		$\ket{0}$ & \-6.392 & 0.697 & \-0.032 & \-0.161 & \-0.032 & 0.697\\
		$\ket{1}$ & \-6.236 & 0.704 & \-0.069 & 0 & 0.069 & \-0.704\\
		$\ket{2}$ & \-2.444 & \-0.069 & \-0.704 & 0 & 0.704 & 0.069\\
		$\ket{3}$ & \-1.005 & \-0.030 & 0.612 & \-0.500 & 0.612 & \-0.030\\
		$\ket{4}$ & 0.577 & 0.114 & 0.354 & 0.851 & 0.354 & 0.114\\
		\br
	\end{tabular}
	\end{indented}
\end{table}

We model the experimental setup of~\cite{hirjibehedin_large_2007} as two reservoirs of noninteracting electrons coupled by a tunnel barrier with an embedded spin. Hence, the Hamiltonian describing the system is given by
\begin{equation}
	H=\sum_{r}H_r+H_\mathsf{spin}+H_\mathsf{tun}.
\end{equation}
Here $H_r=\sum_{\vec k\sigma} \varepsilon_{r\vec k}a_{r\vec k\sigma}^\dagger a_{r\vec k \sigma}$ models the two electrodes as reservoirs of noninteracting electrons with constant density of states $\rho_r$ and electrochemical potential $\mu_r$. The operator $a_{r\vec k\sigma}^\dagger$ creates an electron in lead $r=\mathsf{L,R}$ with momentum $\vec k$ and spin $\sigma$. The local spin is described by
\begin{equation}\label{eq:spin}
	H_\mathsf{spin}=-DS_z^2+E(S_x^2-S_y^2)+g\mu_\mathsf{B}\vec B\cdot\vec S,
\end{equation}
where the $z$ axis is the magnetic easy axis of the atom in its coordination environment. For an $S=2$ iron atom on Cu$_2$N, the best fit to the experimental results in~\cite{hirjibehedin_large_2007} gives an uniaxial anisotropy $D=\unit[1.55]{meV}$, a transverse anisotropy $E=\unit[0.31]{meV}$ and a $g$-factor of $g=2.11$. In \tref{tab:eigenstates}, we summarize the eigenenergies and eigenstates of the spin Hamiltonian for two different choices of the external magnetic field. Finally, the tunnelling Hamiltonian is given by the Appelbaum Hamiltonian~\cite{Appelbaum_s-d_1966}
\begin{equation}
	H_\mathsf{tun}=\sum_{rr'\vec k\vec k'\sigma\sigma'}j_{rr'} a_{r\vec k\sigma}^\dagger\frac{ \bsigma_{\sigma\sigma'}\cdot \vec S}{2}a_{r'\vec k'\sigma'},
\end{equation}
with $\boldsymbol\sigma$ denoting the Pauli matrices, which describes an exchange interaction between the spin of the tunnelling electron and the local spin. We neglect direct tunnelling through the barrier not involving the localized spin as it only gives rise to a bias-independent elastic background to the differential conductance. Interference terms between direct and exchange tunnelling do not appear in the total current and shot noise since the contributions from spin-up and -down electrons cancel out each other for nonmagnetic electrodes. The above model has been studied extensively to describe molecular magnets.
The Kondo effect induced by the transverse anisotropy~\cite{romeike_quantum-tunneling-induced_2006,romeike_kondo-transport_2006,wegewijs_magneto-transport_2007} as well as Berry phase effects~\cite{leuenberger_berry-phase_2006,gonzlez_berry-phase_2007} and the current-induced switching of the molecule spin~\cite{misiorny_quantum_2007} has been discussed.

We parametrize the couplings $j_{rr'}$ through the sum $J=j_\mathsf{LL}+j_\mathsf{RR}$ and the asymmetry $a=(j_\mathsf{LL}-j_\mathsf{RR})/(j_\mathsf{LL}+j_\mathsf{RR})$, i.e., $j_\mathsf{LL}^2=(1+a)^2J^2/4$, $j_\mathsf{RR}^2=(1-a)^2J^2/4$ and $j_\mathsf{LR}^2=j_\mathsf{RL}^2=(1-a^2)J^2/4$. While the couplings $j_\mathsf{LR}$ and $j_\mathsf{RL}$ are responsible for the current through the atom, which may be accompanied with a spin excitation or disexcitation, the couplings $j_\mathsf{LL}$ and $j_\mathsf{RR}$ do not contribute to the current but give rise to a transport-induced relaxation mechanism for the local spin only.

The dynamics of the system is governed by a generalized master equation for the probabilities $P_m$ to find the spin in one of its eigenstates $\ket{m}$ with energy $E_m$,
\begin{equation}\label{eq:master}
	\frac{\rmd P_m}{\rmd t}(t)=\sum_{m'}\int_{-\infty}^t dt' L_{mm'}(t-t')P_{m'}(t') \, .
\end{equation}
In the stationary limit, $P_{m}\equiv P_{m}(t)$ is independent of time and only the time-integrated kernel $L_{mm'}\equiv \int_{-\infty}^0 dt' L_{mm'}(-t')$ is needed.
Its matrix elements are $L_{mm'}=W_{mm'}-\delta_{mm'}W_m$. 
Here, $W_{mm'}$ are the Fermi's golden rule transition rates, 
\begin{equation}\label{eq:rates}
	W_{mm'}=\sum_{rr'\alpha}2\pi|j_{rr'}|^2\rho_r\rho_{r'}|\bra{m}S_\alpha\ket{m'}|^2\zeta(\mu_r-\mu_{r'}-\Delta_{mm'}),
\end{equation}
where $\zeta(x)=x/(1-\rme^{-x/(\kB T)})$, and $\Delta_{mm'}=E_m-E_{m'}$.
The sum runs over the lead indices $r,r'= \mathsf{L},\mathsf{R}$ and the spin directions $\alpha=x,y,z$.
The elements $W_{m}$ follow from $\sum_{m} L_{mm'}=0$ which guarantees the conservation of probability.

To compute the current and current noise, we employ the formalism of full-counting statistics adopted to system that can be described by rate equations~\cite{bagrets_full_2003,braggio_full_2006}.
To this end, we introduce the matrix $W_{mm'}^\chi$ which is obtained from $W_{mm'}$ by multiplying each term in the sum~\eref{eq:rates} with a factor $e^{i\chi}$ if $r=\mathsf{L}, r'=\mathsf{R}$, $e^{-i\chi}$ if $r=\mathsf{R}, r'=\mathsf{L}$, and 0 otherwise, where $\chi$ is called a counting field.
Furthermore, we define $L_{mm'}^\chi=W_{mm'}^\chi-\delta_{mm'}W_m$ (note that $W_m$ does not contain the counting field $\chi$).
The smallest eigenvalue of $L_{mm'}^\chi$ defines the cumulant generating function $S(\chi)$, from which we can obtain the average current $I$ and the current noise $S$ by performing derivatives with respect to the counting field, $I=-\rmi e\left.(\rmd S(\chi)/\rmd\chi)\right|_{\chi=0}$ and $S=(-\rmi e)^2\left.(\rmd^2S(\chi)/\rmd\chi^2)\right|_{\chi=0}$.

Although the full-counting statistics formalism to compute the current and noise is very compact and elegant for the calculation, we introduce, in addition, an equivalent formulation for the average current, that offers a more transparent basis for distinguishing equilibrium from non-equilibrium effect.  
It is easy to show that the average current can also be written as
\begin{equation}
	I=-\rmi e \sum_{m,m'} \left.\frac{\rmd}{\rmd\chi} W_{mm'}^\chi \right|_{\chi=0} P_{m'} \, .
\end{equation}
The derivatives $\rmi (\rmd W_{mm'}^\chi /\rmd\chi)|_{\chi=0}$ are the current rates.
Non-equilibrium effects of the current-voltage characteristics enter via the non-equilibrium probability distribution $P_m$, that is obtained by solving the master equation, Eq.~(\ref{eq:master}).
These non-equilibrium effects would be neglected if one replaced the $P_m$ by an equilibrium probability distribution, $P_m^\mathsf{eq} = \exp(-E_m/\kB T) / \sum_{m'} \exp(-E_{m'}/\kB T)$, i.e., for low temperature $P_0=1$ for the ground state and $P_m=0$ for the excited states $m\neq 0$.

\section{Results}
\begin{figure}
	\includegraphics[width=\textwidth]{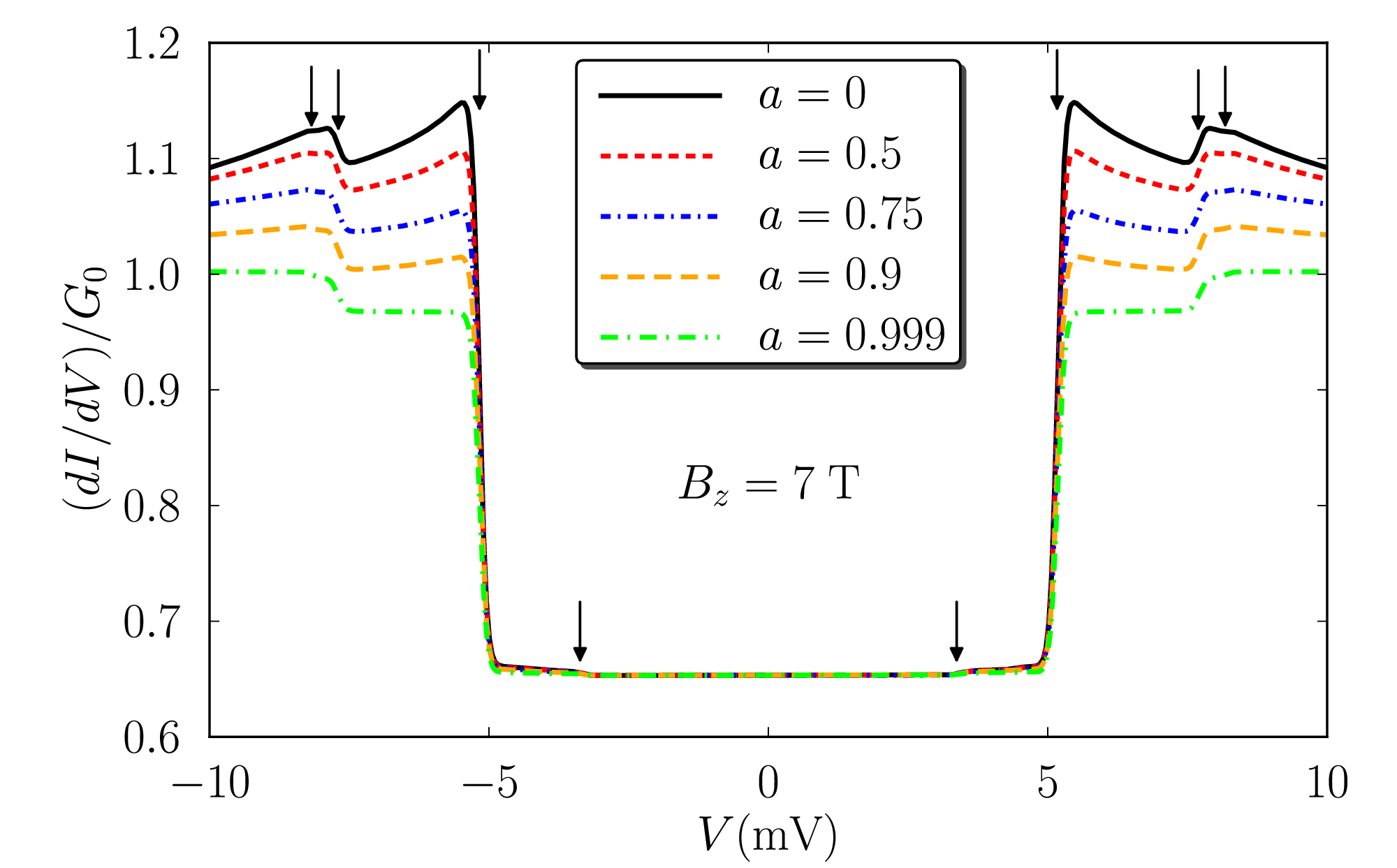}
	\caption{\label{fig:conductance}Differential conductance in units of $G_0$ as a function of bias voltage for different values of the asymmetry parameters $a$. Arrows indicate the position of the excitation energies. Parameters are $B_z=\unit[7]{T}$ and $T=\unit[0.5]{K}$ corresponding to the experimental values of~\cite{hirjibehedin_large_2007}. The corresponding eigenenergies and eigenstates are summarized in \tref{tab:eigenstates}.}
\end{figure}
In the following we discuss the influence of a nonequilibrium spin occupation on the transport properties for the system parameters of the experiment \cite{hirjibehedin_large_2007}. In figure~\ref{fig:conductance}, we show the differential conductance in the presence of a strong magnetic field $B_z=\unit[7]{T}$ along the easy axis for different values of the asymmetry parameter $a$ in the absence of the phenomenological relaxation~\eref{eq:relaxation}, see discussion below. For very large asymmetries, $a\to 1$, there are flat plateaus between the conductance steps. In this limit, the coupling constant $j_{LR}$ for processes that drive the atom state population out of equilibrium is much smaller than $j_{LL}$ for processes that let the system relax to thermal equilibrium with the left electrode. Therefore, as in~\cite{jaklevic_molecular_1966}, nonequilibrium effects are absent, and the resulting conductance curve is identical to the one obtained in~\cite{fernndez-rossier_theory_2009}.
For smaller asymmetries, the situation is different. The height of the conductance steps at the excitation thresholds is increased. Beyond the threshold voltages, the differential conductance shows a slow powerlaw decay towards its value for the equilibrated system again. This overshooting behavior is observed for most of the steps in the experiments of~\cite{hirjibehedin_large_2007}.
While the coupling of the adatom to the substrate is fixed in experiment, the coupling to the STM tip can be controlled by changing the tip-atom distance. This corresponds to changing the total coupling $J$ and thereby the total tunnel current as well as the asymmetry $a$ and thereby the nonequilibrium effects. In a recent experiment using a magnetic STM tip~\cite{loth_controllingstate_2010}, it was confirmed that by decreasing the tip-atom distance and therefore increasing the current through the system, the nonequilibrium effects became more pronounced.

Before we discuss this for the system at hand, we illustrate the mechanism that leads to this conductance behaviour explicitly for the simpler model of a local spin-$1/2$ with Zeeman energy $B$, symmetrically coupled to the electrodes, at zero temperature. Transport takes place by either spin-flip or spin-conserving transitions. The latter contribute to the current as
\begin{equation}
	I_\mathsf{sc}=\pi e|j_\mathsf{LR}|^2\rho_\mathsf{L}\rho_\mathsf{R} eV,
\end{equation}
independent of the probabilities $P_\up$ and $P_\down$ to find the spin in state up and down, respectively.
The differential conductance, measured in units of $G_0=4\pi e^2 S(S+1)|j_\mathsf{LR}|^2\rho_\mathsf{L}\rho_\mathsf{R}$ is $G_\mathsf{sc}=G_0/3$. Therefore, nonequilibrium population of the spin states is only probed by the spin-flip processes. They contribute for $eV\ge B$ as
\begin{equation}\label{eq:currentsf}
	I_\mathsf{sf}=2\pi e|j_\mathsf{LR}|^2\rho_\mathsf{L}\rho_\mathsf{R}\left[(eV-B)P_\up+(eV+B)P_\down\right].
\end{equation}

In equilibrium, only the ground state is occupied, $P_\up=1$ and $P_\down=0$, such that only the first term in~\eref{eq:currentsf} contributes. Hence, the differential conductance $G_\mathsf{sf}^\mathsf{eq}=2G_0/3$ remains constant above threshold.
In the nonequilibrium situation, the occupation probabilities are obtained from the master equation~\eref{eq:master} in the stationary state,
\begin{equation}
	0=\frac{\rmd}{\rmd t}\left(
	\begin{array}{c}
		P_\up\\
		P_\down
	\end{array}
	\right)
	=
	2\pi e|j_\mathsf{LR}|^2\rho_\mathsf{L}\rho_\mathsf{R}\left(
	\begin{array}{cc}
		-(eV-B) & eV+B \\
		eV-B & -(eV+B)
	\end{array}
	\right)
	\left(
	\begin{array}{c}
		P_\up\\
		P_\down
	\end{array}
	\right).
\end{equation}
The solution is $P_\up=1-P_\down=1-\frac{eV-B}{2(eV+B)}$. 
As a consequence, now both terms in \eref{eq:currentsf} contribute, leading to the total conductance (above threshold)
\begin{equation}\label{eq:conductance}
	G=\frac{2}{3}G_0\left(1+\frac{2B^2}{(eV+B)^2}\right).
\end{equation}
In the limit $V\to\infty$, both $P_\up,P_\down\rightarrow 1/2$ and the conductance approaches the equilibrium value $G_0$ with a powerlaw on voltage scale $B$ (although the probability distribution remains highly non-equilibrium).

For the $S=2$ spin of the iron atom with its more complicated spin Hamiltonian, the same mechanism as in the simpler spin-$1/2$ model gives rise to the enhanced conductance in the nonequilibrium situation.
While in equilibrium only the ground state is occupied, $P_0=1$, leading to steps in the differential conductance, in the nonequilibrium case we obtain bias dependent occupations by solving the master equation~\eref{eq:master}, that lead to an overshooting. Again, the conductance decreases above threshold to approach its equilibrium value $G_0$ in the limit of infinite bias voltage. From our analysis it is clear that the nonmonotonic differential conductance is due an {\it increase} of transport enabled by the population of excited states above threshold but close to the step. It is not a signature of the excited spin states carrying less current than the ground state, i.e., a {\it decrease} of the conductance, as has been speculated in~\cite{delgado_spin-transfer_2010}. This conclusion can be experimentally checked by measuring the Fano factor, i.e., the ratio between current noise and average current, $F=S/(eI)$, as we now explain.

\begin{figure}
	\includegraphics[width=\textwidth]{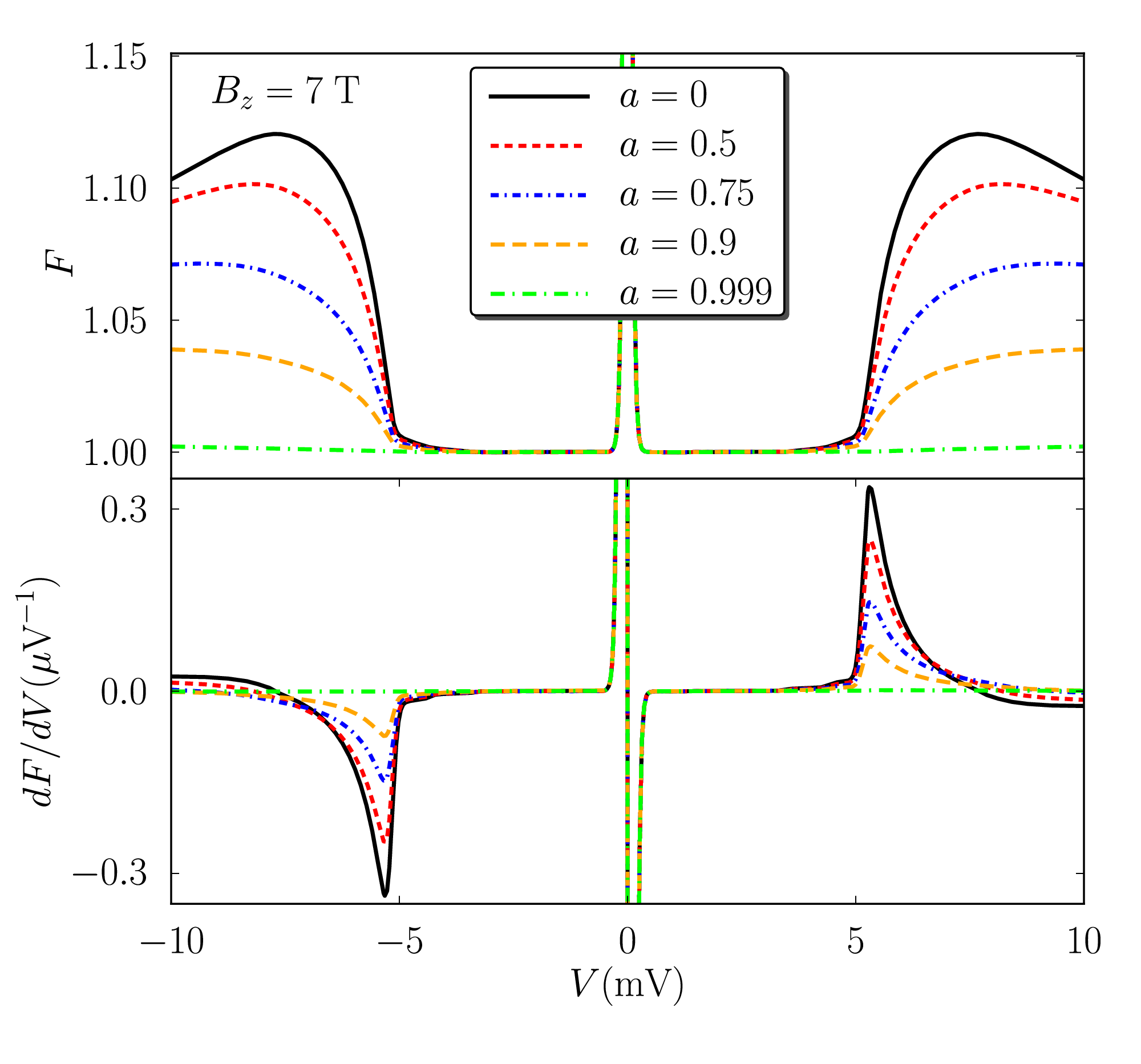}
	\caption{\label{fig:Fano}Fano factor $F$ and $dF/dV$ as a function of bias voltage for different values of the asymmetry parameter $a$. For $eV/\kB T\to 0$ the Fano factor diverges due to thermal noise. Parameters as in figure~\ref{fig:conductance}.}
\end{figure}
The Fano factor $F$ is shown in figure~\ref{fig:Fano} for different values of the asymmetry parameter $a$ not taking into account the relaxation term~\eref{eq:relaxation}, see below.
For $a\to1$, we find $F=1$, i.e., Poissonian behavior, as expected for transport through a normal tunnel barrier.
When a nonequilibrium population of the atom spin states becomes important ($a<1$ and bias voltage exceeding the inelastic threshold), the Fano factor becomes super-Poissonian, reaches a maximum and then slowly drops towards the Poissonian limit for large bias (The transition $\ket{0}\to\ket{1}$ hardly gives rise to super-Poissonian current noise as it is a very weak excitation, cf. figure~\ref{fig:conductance}). The latter behavior is an indicator that the nonmonotonic conductance
is not due to a smaller current contribution from the excited states. If this was the case, we would expect a random telegraph signal with super-Poissonian Fano factor for $V\to\infty$.

The mechanism leading to the super-Poissonian noise for bias voltages above the inelastic thresholds can most easily understood by considering the spin-$1/2$ model again. In this case, the Fano factor above threshold is
\begin{equation}
	F=1+\frac{2B^2}{(eV+B)^2}\cdot\frac{3(eV-B)^2+8(eV-B)B}{3(eV-B)^2+9(eV-B)B+2B^2}.
\end{equation}
Spin-conserving tunnelling processes are stochastically independent of each other and of the spin-flip transitions. They obey Poissonian statistics and can be ignored for the following discussion. Once the inelastic transport channel is open, spin-flip transitions set in. They lead to an alternating sequence of the spin being in the ground and the excited state. In the limit of large bias voltage, the rates for the spin-flip transitions $\up\rightarrow\down$ and $\down\rightarrow\up$ become equal, and the transport statistics becomes Poissonian. For voltages just above threshold, $eV\geq B$, however, the two spin-flip rates differ from each other. As a consequence, we obtain an alternating sequence of a longer and a shorter waiting time, i.e., effectively there is a tendency of two electrons to bunch together, which yields super-Poissonian current noise.

\begin{figure}
	\includegraphics[width=\textwidth]{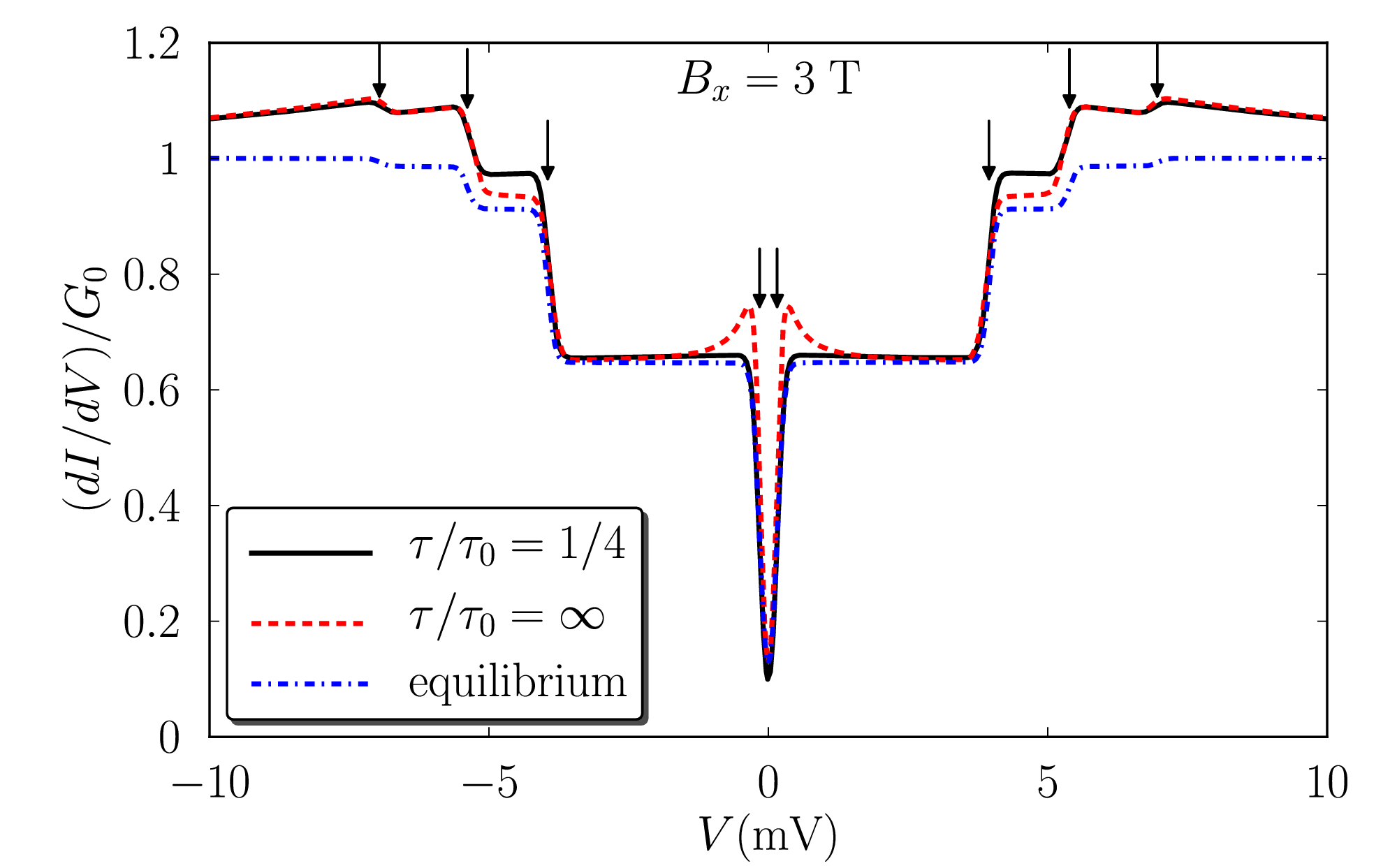}
	\caption{\label{fig:relaxation}Differential conductance taking into account a spin-dependent relaxation mechanism of the form~\eref{eq:relaxation}. The relaxation time is given in units of $\tau_0$ with $\tau_0^{-1}=2\pi DS^2|j_\mathsf{LR}|^2\rho_\mathsf{L}\rho_\mathsf{R}$. Parameters are $B_x=\unit[3]{T}$, $T=\unit[0.5]{K}$, $a=0$. The corresponding eigenenergies and eigenstates are summarized in \tref{tab:eigenstates}.}
\end{figure}

While our theory predicts an overshooting of the differential conductance at {\it all} conductance steps, in the experiment of~\cite{hirjibehedin_large_2007} this feature is absent for the steps associated with the transition between the ground state $\ket{0}$ and the first excited state $\ket{1}$ whenever these steps are pronounced as is the case for a magnetic field along the $x$ axis. This indicates that some relaxation mechanism reduces the occupation of $\ket{1}$. We note that the transition matrix element of $S_z$ between the ground state and the first excited state is large compared to matrix elements of $S_x$ and $S_y$ as well as compared to matrix elements of $S_z$ between the ground state and any other excited state. This observation is not very sensitive to the direction and the strength of the applied magnetic field. Therefore, we make the ad-hoc assumption that there is an additional spin relaxation channel that couples to the $z$-component of the local spin only.

We add to our master equation~\eref{eq:master} the following phenomenological, spin-dependent relaxation rates
\begin{equation}\label{eq:relaxation}
	W_{mm'}^\mathsf{relax}= -\frac{|\bra{m}S_z\ket{m'}|^2}{\tau}\Theta(\Delta_{m'm})
\end{equation}
for $m\neq m'$, where $\Theta(x)$ is the step function and $\tau$ is the time scale for relaxation. The energy dependence in~\eref{eq:relaxation} is not crucial for our conclusions. We therefore choose the simplest possible ansatz that allows relaxation only into states with lower energy. In contrast, the spin matrix elements are crucial as they suppress the nonequilibrium effects for the transition between ground and first excited state while leaving them unaffected for almost all other transitions.

In figure~\ref{fig:relaxation} we plot the differential conductance in the presence of a magnetic field $B_x=\unit[3]{T}$ along the $x$-direction. The first transition is more pronounced than in figure~\ref{fig:conductance} where $B_z=\unit[7]{T}$ along the $z$-direction. In the limit $\tau\to\infty$ we recover the situation discussed above where an overshooting effect can be observed for each conductance step. By choosing a finite value for the relaxation time comparable to the cotunnelling rates exciting the system we can, however, eliminate the overshooting at the first transition while leaving the remaining part of the conductance curve practically unaffected. In the limit $\tau=0$, we recover the equilibrium value for the conductance at each step. These results are not sensitive to the choice of the size and direction of the magnetic field.

The anisotropic relaxation can also explain the absence of conductance steps due to transitions between excited states in the experiment. Such features should be present for a small magnetic field applied in the $z$ direction as in this case the first excited state gets populated significantly at the first conductance step, the excitation energies satisfy $\Delta_{21}<\Delta_{20}$ such that the transition $\ket{1} \to \ket{2}$ occurs before the onset of the transition $\ket{0}\to\ket{2}$ and furthermore the transition matrix elements $\bra{2}S_\alpha\ket{1}$ do not all vanish. However, as the relaxation prevents the system from populating the first excited state, these additional conductance features vanish together with the overshooting at the first step.

\section{Conclusions}
We investigated the nonequilibrium effects in transport through a single iron atom. With our model, we were able to explain the nonmonotonic features of the differential conductance observed experimentally in~\cite{hirjibehedin_large_2007}. Furthermore, we noted a striking absence of this nonmonotonicity at certain conductance steps which can be explained by the presence of an anisotropic spin relaxation channel. The anisotropy~\cite{leuenberger_spin_2000} points at the importance of spin-orbit coupling in this process, in addition to the splitting of the multiplet. In addition, we predicted the occurrence of super-Poissonian current noise as a consequence of the nonequilibrium spin occupations probabilities. In conclusion, for a full understanding of inelastic tunnelling spectroscopy, it is crucial to account for nonequilibrium populations of the atom states established by the competition of transport and anisotropic relaxation.

\ack
We acknowledge financial support from DFG via SFB 491.

\section*{References}


%

\end{document}